\documentclass[conference]{IEEEtran}
\IEEEoverridecommandlockouts

\usepackage{booktabs}
\usepackage{cite}
\usepackage{amsmath,amssymb,amsfonts}
\usepackage{algorithmic}
\usepackage{graphicx}
\usepackage{textcomp}
\usepackage{xcolor}
\usepackage{tabularx}
\def\BibTeX{{\rm B\kern-.05em{\sc i\kern-.025em b}\kern-.08em
    T\kern-.1667em\lower.7ex\hbox{E}\kern-.125emX}}

\usepackage{xurl}
\usepackage{hyperref}
\usepackage{todonotes}

\usepackage{listings}

\usepackage{tikz}
\usetikzlibrary{shapes.geometric, arrows.meta, positioning}

\definecolor{codegray}{rgb}{0.95,0.95,0.95} 
\definecolor{codeblue}{rgb}{0,0,0.6}
\definecolor{codegreen}{rgb}{0,0.5,0}
\definecolor{codered}{rgb}{0.6,0,0}

\newif\ifshowreview
\showreviewtrue
\DeclareRobustCommand{\review}[1][]{%
  \ifshowreview
    \textcolor{blue}{#1\if\relax\detokenize{#1}\relax\else\,\fi}%
  \else
    #1%
  \fi}

\setuptodonotes{inline, color=orange!30, size=\footnotesize}

\begin{document}

\title{Enabling Dynamic Resource Scaling in Quantum Circuit Simulations}


\author{\IEEEauthorblockN{1\textsuperscript{st} Íñigo Aréjula-Aísa}
\IEEEauthorblockA{\textit{Barcelona Supercomputing Center} \\
\textit{(BSC)}\\
0009-0006-7354-0915}
\and
\IEEEauthorblockN{2\textsuperscript{nd} Sergio Iserte}
\IEEEauthorblockA{\textit{Barcelona Supercomputing Center} \\
\textit{(BSC)}\\
0000-0003-3654-7924}
\and
\IEEEauthorblockN{3\textsuperscript{rd} Petter Sandås}
\IEEEauthorblockA{\textit{Barcelona Supercomputing Center} \\
\textit{(BSC)}\\
0009-0009-2028-7719}
\and
\IEEEauthorblockN{3\textsuperscript{th} Ricard. S. Raigada-García}
\IEEEauthorblockA{\textit{Universitat Politècnica de Catalunya} \\
\textit{(UPC)}\\
Barcelona, Spain \\
0009-0009-2028-7719}
\and
\IEEEauthorblockN{4\textsuperscript{th} Oliver Brown}
\IEEEauthorblockA{\textit{Edinburgh Parallel Computing Centre} \\
\textit{(EPCC)}\\
Edinburgh, UK \\
email address or ORCID}
\and
\IEEEauthorblockN{5\textsuperscript{th} James Richings}
\IEEEauthorblockA{\textit{Edinburgh Parallel Computing Centre} \\
\textit{(EPCC)}\\
Edinburgh, UK \\
email address or ORCID}
\and
\IEEEauthorblockN{8\textsuperscript{th} Antonio J. Peña}
\IEEEauthorblockA{\textit{Barcelona Supercomputing Center} \\
\textit{(BSC)}\\
Barcelona, Spain \\
0000-0002-3575-4617}}

\maketitle

\begin{abstract}
Studies of quantum algorithms often simulate sequences of circuits whose number of qubits changes from one circuit to the next, such as instances of increasing size of the same problem. Since each additional qubit doubles the state-vector, every circuit needs a different number of nodes, which a fixed allocation on a shared High-Performance Computing (HPC) system cannot provide: a job sized for the largest circuit leaves most of its nodes idle, and one job per circuit adds scheduler interactions and queue waits. We make the QuEST state-vector simulator malleable with the Dynamic Management of Resources (DMR) framework, so that a single job resizes between circuits. Because no state-vector exists between circuits, reconfigurations require no data redistribution and no changes to the simulation kernels, and a checkpoint holding the index of the next circuit is enough to resume. A right-sizing policy derives the number of nodes from the number of qubits, and the \texttt{circuit\_executor} API hides reconfigurations from the user. 
We evaluate the reconfiguration overhead on MareNostrum~5 and show that a single job can run a batch of circuits of different sizes, adapting its allocation to each of them with negligible overhead.
\end{abstract}


\section{Introduction}\label{sec:intro}
Quantum computing (QC) is expected to speed up certain problems in chemistry, materials science, and optimization~\cite{montanaro_2016}. However, current and near-term quantum processors are limited by noise, small qubit counts, and shallow circuit depths, a regime known as Noisy Intermediate-Scale Quantum (NISQ)~\cite{Preskill2018NISQ}.

Until fault-tolerant hardware becomes available, quantum circuit simulators allow researchers to design, benchmark, and refine quantum algorithms on classical computers. Simulations show how algorithms scale, how robust they are to noise, and how the constraints of real hardware affect them. Results from independent simulators are also cross-checked to validate complex circuits~\cite{jamadagni_benchmarking_2024}.

These studies often require many simulations, for instance to sweep parameters or noise models, and the cost of each simulation grows exponentially with the number of qubits. Simulators therefore rely on multithreading, distributed memory, and GPU acceleration, and run on platforms ranging from laptops to supercomputers.

QuEST (Quantum Exact Simulation Toolkit)~\cite{jones_quest_2019} is an open-source simulator of this kind. It simulates both state-vectors and density matrices, and thus both ideal and noisy circuits. It combines OpenMP multithreading, GPU acceleration with CUDA, and distribution with the Message Passing Interface (MPI), which allows it to simulate large circuits on supercomputers.

Moreover, the size of the simulated circuits often changes and later circuits can depend on the results of earlier ones. A common pattern is to optimize the parameters of a variational algorithm on small instances and reuse them on larger ones. For instance, the scaling of the Quantum Approximate Optimization Algorithm (QAOA)~\cite{farhi_qaoa_2014} on the low autocorrelation binary sequences (LABS) problem has been studied with exact state-vector simulations of 24 to 40 qubits on up to 1,024 GPUs per simulation, using a single set of parameters obtained by optimizing on small instances and rescaled to every problem size~\cite{shaydulin_labs_2024}. Similarly, QAOA parameters optimized on small graphs can be transferred to much larger ones~\cite{galda_transfer_2023}. Since each additional qubit doubles the size of the state-vector, the resources needed by a distributed simulation change with every circuit in the sequence. Static allocations on shared High-Performance Computing (HPC) systems fit such sequences poorly: a job sized for the largest circuit keeps most of its nodes idle while the smaller circuits run, whereas one job per circuit multiplies the interactions with the scheduler and exposes every circuit to its own queue wait. This becomes worse when the system operates under restrictions, for instance by limiting the number of nodes per user or allowing only one running job at a time, since these jobs then cannot run concurrently and their queue waits add up one after another.

Malleability, the ability of applications to adjust their resources at runtime, addresses this mismatch by allowing a simulation to expand or shrink its footprint as the sequence progresses. Malleable jobs can improve system utilization and reduce wait times~\cite{iserte_mpi_2025}; a malleable job waits in the queue once, at submission, and although later expansions still depend on the availability of free nodes, shrinkages return idle nodes to the system. However, QuEST, like other distributed simulators, runs on an allocation fixed at launch.

To close this gap, we integrate QuEST with the Dynamic Management of Resources (DMR) framework~\cite{10.1007/978-3-031-85697-6_19}, which allows MPI applications to expand and shrink at runtime in coordination with the resource manager. We present a case study in which a single QuEST job runs a sequence of circuits of different sizes and adapts its number of nodes to each of them. We measure the reconfiguration overhead and discuss the usability benefits of this approach.

The contributions of this paper are: (i) the design and implementation of a malleable version of QuEST that reconfigures at circuit boundaries, using the checkpoint-restart mode of DMR; (ii) a right-sizing policy that maps the number of qubits of each circuit to a number of nodes; (iii) the \texttt{circuit\_executor} API, which hides reconfigurations from the user; and (iv) an evaluation of the reconfiguration overhead on MareNostrum~5.

The rest of the paper is organized as follows. Section~\ref{sec:background} provides background on QuEST and DMR. Section~\ref{sec:related} reviews related work on malleability, distributed quantum simulation, and resource management for HPC-QC integration. Section~\ref{sec:methodology} describes the design of the integration, Section~\ref{sec:evaluation} evaluates it, Section~\ref{sec:discussion} discusses our findings and their limitations, and Section~\ref{sec:conclusion} concludes the paper.

\section{Background}\label{sec:background}
This section provides background on the two main components of our study: QuEST, a high-performance quantum circuit simulator, and DMR, a framework for enabling malleability in MPI applications.

\subsection{QuEST}\label{subsec:quest}

QuEST~\cite{jones_quest_2019} is an open-source C/C++ library for the exact simulation of quantum circuits on state-vectors and density matrices. A state-vector of $n$ qubits holds $2^n$ complex amplitudes, 16~bytes each in double precision, and a density matrix of $n$ qubits is stored as a vector of $2^{2n}$ amplitudes. 
QuEST provides different back-end algorithms~\cite{jones_distributed_2023}, and combines OpenMP multithreading, GPU acceleration, and memory distribution with MPI.

In distributed mode, the $2^n$ amplitudes are split evenly among $P = 2^k$ MPI processes, each holding a contiguous block of $2^{n-k}$ amplitudes (Fig.~\ref{fig:quest-dist}). Hence, $P$ must be a power of two and cannot exceed $2^n$, since every process must hold at least one amplitude. Gates acting only on the $n-k$ low-order qubits are local and need no MPI communication, whereas the $k$ high-order qubits index the process that holds an amplitude, so a gate on one of them pairs processes that exchange amplitudes. For this exchange, each process allocates a communication buffer of the same size as its block, which doubles the memory per process. 

\begin{figure}[t]
\centering
\resizebox{\columnwidth}{!}{%
\begin{tikzpicture}[
    font=\footnotesize,
    blk/.style={draw, minimum height=7mm, text width=17mm, align=center, inner sep=1pt},
    buf/.style={draw, dashed, minimum height=4mm, text width=17mm, align=center, inner sep=1pt},
    pair/.style={{Stealth[length=1.4mm]}-{Stealth[length=1.4mm]}},
]
\foreach \r in {0,1,2,3} {
  \node[blk] (b\r) at (\r*2.05,0) {process $\r$\\$2^{n-2}$ amplitudes};
  \node[buf, anchor=north] (c\r) at ([yshift=-0.8mm]b\r.south) {buffer};
}
\draw[pair] (b0.north) to[bend left=25] node[above] {gate on $q_{n-1}$} (b2.north);
\draw[pair] (c1.south) to[bend right=25] (c3.south);
\node[anchor=north, align=center] at ([yshift=-1mm]current bounding box.south)
  {local qubits $q_0,\dots,q_{n-3}$; \ $q_{n-2}, q_{n-1}$ select the process};
\end{tikzpicture}}
\caption{Distribution of an $n$-qubit state-vector over $P=4$ MPI processes in QuEST ($k=2$). Each process stores a block of $2^{n-2}$ amplitudes and an equally sized communication buffer. Gates on local qubits need no communication; a gate on global qubit $q_{n-1}$ pairs processes $0\leftrightarrow2$ and $1\leftrightarrow3$, which exchange amplitudes through their buffers via MPI.}
\label{fig:quest-dist}
\end{figure}

\subsection{DMR}\label{subsec:dmr}
The \textbf{Dynamic Management of Resources (DMR)} framework\footnote{Pronounced ``dimmer.''} is a middleware layer that enables process malleability~\cite{feitelson_packing_1996} and runtime resource management for MPI applications in HPC environments. DMR acts as an orchestration layer connecting scientific applications, the MPI runtime (e.g., Open MPI/PRRTE (PMIx Reference Runtime Environment)~\cite{gabriel_2004,prrte,castain_pmix_2018}), performance monitoring tools (e.g., TALP~\cite{lopez_talp_2021}), and the underlying Resource Management System (RMS), e.g., Slurm~\cite{yoo_slurm_2003}. Its primary goal is to allow running jobs to dynamically acquire, release, or reconfigure compute resources at runtime, enabling applications to transparently expand or shrink their resource footprint, thereby improving system utilization and responsiveness.

Internally, DMR implements a state-driven architecture exposed to developers through a core API~\cite{iserte_dmrlib_2020}:
\begin{itemize}
    \item \texttt{dmr\_init}: Initializes the dynamic environment, establishes communication with the RMS, and sets up internal data structures.
    \item \texttt{dmr\_check}: Called at synchronization points to indicate readiness for resizing; it evaluates policies and system feedback to determine whether reconfiguration is needed.
    \item \texttt{dmr\_reconfigure}: Performs the selected reconfiguration and coordinates with the MPI runtime to spawn or terminate processes as required.
    \item \texttt{dmr\_finalize}: Cleans up the management layer and releases allocated resources.
\end{itemize}

A key advancement is the introduction of a \textit{non-invasive methodology} tailored for production HPC clusters. While earlier approaches relied on customized or invasive scheduler modifications (e.g., Slurm4DMR) in controlled environments~\cite{10.1007/978-3-031-85697-6_19}, the new DMR@Jobs feature enables malleable applications to interact directly with standard, unmodified RMS deployments through user-level API calls~\cite{sandas_dmrv2_2026}. This approach orchestrates MPI processes across multiple independent jobs to achieve expansion and shrinkage without requiring administrative privileges. DMR also provides a unified programming model that supports both \textbf{in-memory data redistribution} via MPI and \textbf{checkpoint-restart (C/R)} mechanisms, adapting to diverse scientific codes and existing application-level checkpointing logic.

The framework supports sophisticated reconfiguration strategies which allow applications to continue executing while resource requests are pending, so DMR hides resource acquisition latency in heavily contended production systems. This makes it a robust solution for bridging the gap between experimental malleability prototypes and operational supercomputing environments. 


\section{Related Work}\label{sec:related}
This section reviews work on malleability in HPC, on distributed state-vector simulation, and on resource management for HPC-QC integration, and positions our contribution with respect to them.

\paragraph{Malleability in HPC} Feitelson and Rudolph classify parallel jobs by when and by whom their resources are decided~\cite{feitelson_rudolph_1996}. Rigid and moldable jobs keep the resources assigned at launch, evolving jobs request changes initiated by the application, and malleable jobs adapt to changes decided by the system. Following common usage in MPI research, we use \emph{malleability} for any job that changes its resources at runtime, whether the change is triggered by the application or by the RMS. Malleability can raise system throughput and resource utilization and shorten waiting times, yet it remains rare in production systems~\cite{tarraf_malleability_2024}. It requires support from the RMS, the MPI runtime, and the application, which must redistribute its data among a changing set of processes. At the runtime level, MPI provides dynamic process management to spawn processes.
Malleability has traditionally been evaluated under dynamic workloads of scientific applications~\cite{iserte_resource_2025} and replayed production conditions~\cite{iserte_mpi_2025}.
In this work, we apply malleability to an MPI distributed quantum simulator and reconfigure it between circuits, where no state-vector must be redistributed.

\paragraph{Distributed state-vector simulation} Since a state-vector of $n$ qubits holds $2^n$ amplitudes, large circuits require distributing the state over many nodes. JUQCS~\cite{deraedt_juqcs_2019}, Intel-QS~\cite{guerreschi_intelqs_2020}, QuEST~\cite{jones_quest_2019}, SV-Sim~\cite{li_svsim_2021}, mpiQulacs~\cite{imamura_mpiqulacs_2022}, Qiskit Aer~\cite{doi_cacheblocking_2020}, and cuStateVec~\cite{bayraktar_cuquantum_2023} partition the state-vector among processes or GPUs and exchange amplitudes for gates on non-local qubits; Jamadagni~\emph{et al.}\ benchmark many of these packages on an HPC cluster~\cite{jamadagni_benchmarking_2024}. For QuEST, Adamski~\emph{et al.}\ show that node type and CPU frequency trade runtime against energy and the compute units charged to the user~\cite{adamski_energy_2023}, which illustrates that the best allocation depends on the circuit. All these simulators run on an allocation fixed at launch. Intel-QS can divide its allocation into groups that simulate a pool of related circuits in parallel~\cite{guerreschi_intelqs_2020}, which also targets batches of circuits, but the partition holds for the whole execution. To the best of our knowledge, no distributed simulator adapts its allocation to the size of each circuit at runtime.

\paragraph{Resource management for HPC-QC integration} The integration of quantum processors and simulators into HPC environments has motivated new programming tools~\cite{Elsharkawy2023IntegrationQAcc} and resource-management strategies. Rocco~\emph{et al.}\ use malleability and workflows to release classical resources while computation is offloaded to a quantum processor and to reclaim them afterwards~\cite{rocco_dynamic_2025}, and Cipollini~\emph{et al.}\ compare strategies to share a quantum processor among hybrid applications~\cite{cipollini2026waysshareqpuscheduling}. Closed-loop electronic-structure calculations already couple a quantum processor with a full supercomputer~\cite{shirakawa2025closedloopcalculationselectronicstructure}, and circuit-cutting workflows split a circuit into many smaller subcircuits that are orchestrated on HPC resources~\cite{tejedor2025,garciaraigada2026}. These works manage the classical resources around quantum stages. In contrast, we make the simulation stage itself malleable, so that it can be resized like any other classical stage of such workflows.

\paragraph{Our position} This work lies at the intersection of the three lines above. It applies the malleability support of DMR to a distributed quantum simulator, which, to the best of our knowledge, has not been done before. The partitioning of QuEST remains unchanged; only the allocation it runs on changes between circuits. Resizes are driven by the application itself: the number of qubits of each circuit determines the allocation, and circuit boundaries provide reconfiguration points where no distributed state must be redistributed. The resulting malleable simulator can serve as a resizable stage in the HPC-QC workflows discussed above.

\section{Methodology}\label{sec:methodology}
This section describes how we have designed and implemented the integration of
DMR\footnote{\url{https://zenodo.org/records/17831722}} with QuEST\footnote{\url{https://quest.qtechtheory.org}} to enable
MPI malleability in quantum circuit simulations. We describe where and how reconfigurations take place (Section~\ref{subsec:design}), the policy that decides the size of the job (Section~\ref{subsec:resize}), and the programming interface exposed to users (Section~\ref{subsec:api}).

In this regard, we aim to: (i) adapt the number of nodes of a single job to each of its circuits, (ii) leave the simulation kernels of QuEST untouched, confining the changes to register creation, and (iii) hide the reconfiguration logic from the user.

\subsection{Reconfiguration at Circuit Boundaries}\label{subsec:design}
QuEST is designed with modularity in mind: its distributed-memory parallelism is provided by a separate MPI module that can be enabled or disabled at compile time. Following the same approach, we add a malleability module on top of the distributed-memory layer that interfaces with DMR, so that the core simulation logic is not modified.

The central design decision is to reconfigure only \emph{between} circuits, never during their execution. Each circuit allocates its quantum register, applies its gates, and destroys the register. At a circuit boundary there is therefore no distributed state-vector alive, and the only state that must survive a reconfiguration is the progress of the batch and the results of completed circuits, which each circuit writes to persistent storage and subsequent circuits can read as their input. This has two consequences:

\begin{enumerate}
  \item No state-vector redistribution is needed, which would otherwise involve handling up to $2^n$ amplitudes of data. Instead, the execution is relaunched on an expansion or a shrinkage; DMR respawns the MPI processes and the QuEST simulation restarts from the beginning of its entry point.
  
  \item The checkpoint only contains the index of the next circuit to run. On restart, circuits before this index are skipped and execution resumes at the correct position.
\end{enumerate}

In QuEST, the most straightforward malleability point, i.e., the reconfiguration point where the MPI ranks are synchronized, is the creation of the quantum register. \texttt{createQureg} and its variants invoke the malleability module before allocating the register, which is the earliest moment at which the size of the circuit, $n$, is known and the previous register has already been released. If a resize is performed, the application restarts, skips the completed circuits, and re-enters the current one; this time the register creation finds the correct number of nodes and proceeds. 

After a resize, the only repeated work is the part of the current circuit that precedes the creation of its register. Any action with side effects in this part, such as writing output or seeding a random number generator, is executed again after the restart. Therefore, circuits must create their register before producing any output, and all the registers of a circuit must have the same size, since every register creation is a reconfiguration point. Because gates are always applied on a stable allocation, malleability cannot interfere with the simulation itself.

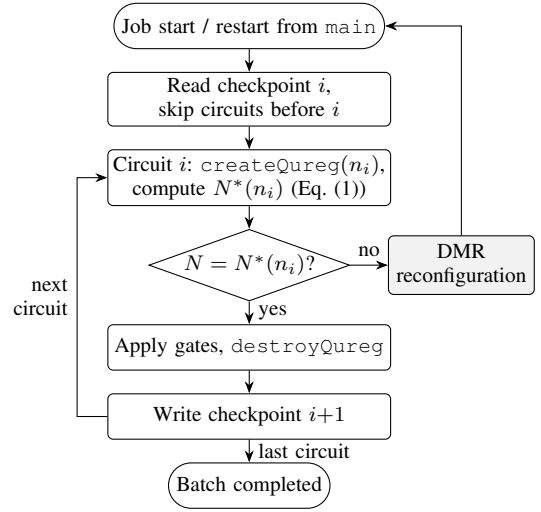
\begin{figure}[t]
\centering
\begin{tikzpicture}[
    font=\footnotesize,
    node distance=3mm,
    box/.style={draw, rounded corners=2pt, align=center, text width=3.6cm, minimum height=6mm, inner sep=2pt},
    term/.style={draw, rounded corners=3mm, align=center, minimum height=6mm, inner sep=3pt},
    dmr/.style={draw, rounded corners=2pt, align=center, fill=codegray, inner sep=3pt},
    dec/.style={draw, diamond, aspect=2.8, align=center, inner sep=0.5pt},
    arr/.style={-{Stealth[length=1.6mm]}},
]
\node[term] (start) {Job start / restart from \texttt{main}};
\node[box, below=of start] (ckpt) {Read checkpoint $i$,\\skip circuits before $i$};
\node[box, below=of ckpt] (create) {Circuit $i$: \texttt{createQureg}$(n_i)$,\\compute $N^{*}(n_i)$ (Eq.~\eqref{eq:rightsize})};
\node[dec, below=of create] (dec) {$N = N^{*}(n_i)$?};
\node[box, below=of dec] (run) {Apply gates, \texttt{destroyQureg}};
\node[box, below=of run] (save) {Write checkpoint $i{+}1$};
\node[term, below=of save] (end) {Batch completed};
\node[dmr, right=5mm of dec] (resize) {DMR \\ reconfiguration};

\draw[arr] (start) -- (ckpt);
\draw[arr] (ckpt) -- (create);
\draw[arr] (create) -- (dec);
\draw[arr] (dec) -- node[right] {yes} (run);
\draw[arr] (run) -- (save);
\draw[arr] (save) -- node[right] {last circuit} (end);
\draw[arr] (dec) -- node[above] {no} (resize);
\draw[arr] (resize) |- (start);
\draw[arr] (save.west) -- ++(-4mm,0) |- node[pos=0.25, left, align=right] {next\\circuit} (create.west);
\end{tikzpicture}
\caption{Execution flow of a malleable batch. Reconfigurations only take place at register creation, when no state-vector is alive; after a resize, the job restarts and resumes from the checkpointed circuit index.}
\label{fig:flow}
\end{figure}

Fig.~\ref{fig:flow} summarizes this execution flow.

\subsection{Automatic Node Right-Sizing}\label{subsec:resize}
At each reconfiguration point, a DMR reconfiguration policy designed for QuEST computes the target number of nodes for a register of $n$ qubits as
\begin{equation}
    N^{*}(n) = \min\left(\max\left(2^{\,n-L},\, 1\right),\, N_{\max}\right),
    \label{eq:rightsize}
\end{equation}
where $L$ is the number of local qubits per node, so that each node holds $2^{L}$ amplitudes of the state-vector, and $N_{\max}$ is the maximum number of nodes the job is allowed to hold. 
Registers with $n \leq L$ fit in a single node, and each additional qubit beyond $L$ doubles the number of nodes until $N_{\max}$ is reached.
Since we run one MPI process per node, in the notation of Section~\ref{subsec:quest} the policy sets $k = n - L$, so that each process keeps $n-k = L$ local qubits, unless the allocation is capped at one node or at $N_{\max}$.
We set $L = 26$, the threshold that QuEST~v4 itself uses to decide whether \texttt{createQureg} distributes a register: it does so only if each node keeps at least $2^{26}$ amplitudes.
If the current node count $N$ equals $N^{*}(n)$, execution continues normally. If $N < N^{*}(n)$, the policy requests an expansion before running the circuit; if $N > N^{*}(n)$, it requests a shrinkage, and the released nodes are returned to the RMS, where they become available to other jobs.
Hence, $N^{*}(n)$ is the largest allocation over which QuEST still distributes an $n$-qubit register; on a larger one, \texttt{createQureg} would not distribute it and the additional nodes would do no useful work.

The policy is evaluated synchronously: the executor waits for the outcome of the reconfiguration before running the circuit, instead of overlapping the wait with computation. The outcome is a request rather than a command, since DMR evaluates it against the state of the system; if it is not granted, the executor waits and retries the request until it is.

\subsection{Programming Interface}\label{subsec:api}
Using the mechanism above directly would require the user to guard each circuit against re-execution after a restart and to manage the DMR environment explicitly. To hide these details, we introduce the \texttt{circuit\_executor}, which manages the execution of a batch of circuits. Users register each circuit as a function, and the executor runs them in order, maintains the checkpoint, and wraps the initialization and finalization of DMR. A circuit is a plain C function with no arguments and no return value (\texttt{void (*)(void)}) that creates its register, applies its gates, reports its results, and destroys the register, so that it is self-contained and can be re-executed from the start after a restart.

Listing~\ref{lst:executor} shows a batch of three circuits of different sizes. Compared with a non-malleable QuEST program, the changes are limited to wrapping each circuit in a function, registering it, and calling the executor.

\begin{figure}
\begin{lstlisting}[language=C, caption=Running a set of circuits with the \texttt{circuit\_executor} API., label=lst:executor, captionpos=b]
initQuESTEnv();
circuit_executor_init(argc, argv);

circuit_executor_add(circuit_32q);
circuit_executor_add(circuit_16q);
circuit_executor_add(circuit_8q);

circuit_executor_run();

circuit_executor_destroy();
finalizeQuESTEnv();
\end{lstlisting}
\end{figure}

Because the whole batch is known before execution starts, the executor runs circuits in registration order; no reordering is performed.

\section{Evaluation}\label{sec:evaluation}
We first measure the cost of a reconfiguration in isolation (Section~\ref{subsec:overhead}) and then run a batch of circuits with the \texttt{circuit\_executor} (Section~\ref{subsec:batch}).

All the experiments are executed on MareNostrum~5 (MN5), a pre-exascale system integrated into the EuroHPC-JU infrastructure\footnote{\url{https://bsc.es/marenostrum/marenostrum-5}}.
Specifically, jobs are submitted to its general-purpose partition (GPP)\footnote{\url{https://www.bsc.es/supportkc/docs/MareNostrum5/overview/\#marenostrum-5-gpp-general-purpose-partition}}, which comprises 6,192 Intel Sapphire Rapids-based nodes, each equipped with two Intel Xeon Platinum 8480+ processors (56 cores at 2~GHz, 112 cores total) and 256~GB of DDR5 memory, interconnected by a 100~Gbit/s ConnectX-7 NDR200 InfiniBand network. 

The core software stack relies on Open MPI~5.1.0a1, PRRTE~5.0.0a1, OpenPMIX~7.0.0a1, DLB~3.5.0, and DMR~2.0.2. QuEST~v4.2.0\footnote{\url{https://github.com/QuEST-Kit/QuEST}}, extended with the malleability module designed in this work, is built in Release mode in double precision, with distribution (MPI) and multithreading (OpenMP) enabled. Each node runs a single MPI process with 112 OpenMP threads, and nodes are allocated exclusively. To avoid the waiting times of the production queue and make the experiments easier to reproduce, they run under Slurm4DMR, a malleability-enabled, user-level instance of Slurm deployed inside a regular MN5 job.

\subsection{Malleability Overhead}\label{subsec:overhead}
\begin{figure}[t]
\centering
\includegraphics[width=\columnwidth]{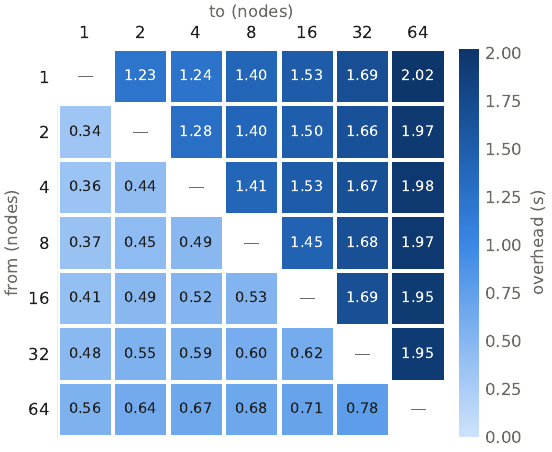}
\caption{Reconfiguration overhead on MN5 (mean of three repetitions). Rows give the source and columns the target number of nodes: expansions lie above the diagonal and shrinkages below it.}
\label{fig:reconf-matrix}
\end{figure}

Before running a batch of circuits, we measure the cost of a reconfiguration in isolation. A minimal QuEST application performs every transition between the power-of-two node counts from 1 to 64. For each pair $(a,b)$, a separate job is launched on $a$ nodes and expands or shrinks to $b$ nodes.
Each transition is repeated three times, and Fig.~\ref{fig:reconf-matrix} reports the mean overhead.

The measured time covers the whole transition: DMR evaluates the action (expand, shrink, or none); in an expansion, the new nodes are acquired from Slurm and added to the PRRTE distributed virtual machine, whereas in a shrinkage the released nodes are returned to Slurm (and asynchronously removed from PRRTE); the processes are respawned on the target nodes; and the MPI, DMR, and QuEST environments are initialized in the new processes. The respawn uses \texttt{MPI\_Comm\_spawn}, whose cost depends on the number of created processes~\cite{iker_spawn}. No quantum register is allocated during this interval, since a circuit creates its register only after the reconfiguration completes.

Expansions take between 1.23 and 2.02~s, and their cost is set by the target size: about 1.25~s to reach 2 or 4 nodes, 1.45~s to 8 or 16, 1.7~s to 32, and 2.0~s to reach 64 nodes.
Shrinkages take between 0.34 and 0.78~s and grow slowly with both sizes: shrinking to one node takes from 0.34~s (from two nodes) to 0.56~s (from 64), and shrinking from 64 nodes takes from 0.56~s (to one node) to 0.78~s (to 32).

The gap between both directions comes from the steps that only expansions perform synchronously. 
Before spawning processes, an expansion must obtain the new nodes from Slurm and extend the PRRTE distributed virtual machine to them, whereas a shrinkage removes nodes as a background operation because they do not participate in the spawning. 

\subsection{Batch Execution}\label{subsec:batch}
We run a batch of circuits with the \texttt{circuit\_executor} API. Since reconfigurations take place only between circuits, every circuit runs entirely on a fixed allocation: from the moment its register is created, through the application of its gates, until its result is read, the circuit executes exactly as it would in a non-malleable run. Malleability only determines the number of nodes on which each circuit starts.

Each circuit applies the Quantum Fourier Transform (QFT) to an $n$-qubit register. The register is allocated with \texttt{createQureg}, initialized to $|0\rangle^{\otimes n}$ with \texttt{initZeroState}, and transformed by a sequence of Hadamard gates and controlled phase shifts (\texttt{apply\allowbreak Hadamard} and \texttt{apply\allowbreak TwoQubit\allowbreak PhaseShift}). An $n$-qubit QFT applies $n$ Hadamard gates and $n(n-1)/2$ controlled phase shifts, and each gate updates every amplitude of the local block. Finally, the probability of the basis state $|0\cdots0\rangle$ is computed with \texttt{calcProbOf\allowbreak BasisState}.

The batch consists of seven circuits of 28, 30, 24, 36, 31, 32, and 34 qubits, executed in this order. With $L=26$ and $N_{\max}=64$, Eq.~\eqref{eq:rightsize} assigns them 4, 16, 1, 64, 32, 64, and 64 nodes, respectively (see the first and second columns of Table~\ref{tab:walltimes}). Keeping $2^{26}$ amplitudes per node, the 34- and 36-qubit circuits would need 256 and 1,024 nodes; they are therefore capped at $N_{\max}$, and each node holds $2^{28}$ and $2^{30}$ amplitudes, respectively. We chose this order to alternate register sizes, so that consecutive circuits require different allocations and a single run includes both expansions and shrinkages, while the last circuit covers the case in which no reconfiguration is needed.

Users do not need to size the job for its circuits: the job can be submitted requesting a single node, and the executor then adapts its resources. Before the first circuit (28 qubits), the executor expands the job to 4 nodes. It then expands to 16 nodes for the 30-qubit circuit, shrinks to a single node for the 24-qubit circuit, expands to 64 nodes for the 36-qubit circuit, shrinks to 32 nodes for the 31-qubit circuit, and expands back to 64 nodes for the 32-qubit circuit. The 34-qubit circuit runs on the same 64 nodes without reconfiguring, since DMR is only invoked when the right-sizing policy selects a different number of nodes. The batch therefore performs six reconfigurations, four expansions and two shrinkages, which are listed as transitions in the fourth column of Table~\ref{tab:walltimes}. Submitting the job directly on 4 nodes would avoid the first expansion, but requires the user to know the size of the first circuit and the right-sizing policy in advance.

Without malleability, this workload would run either as a static 64-node job, with up to 63 idle nodes during the smaller circuits, or as seven independent jobs, each with its own queue wait. With the \texttt{circuit\_executor}, the entire batch is handled within a single job that dynamically adjusts its footprint, reducing the total number of scheduler interactions and the associated queue wait time of each job. Notice that when expanding in the malleable scenario, the requests are also affected by the queue wait time, although these requests are smaller, since only the missing nodes are requested.

The third column of Table~\ref{tab:walltimes} presents the execution time of the circuits, which is driven by the number of amplitudes per node (computational load) and the number of involved global qubits (internode communication). In particular, in direct qubit-node relationships where the number of amplitudes is proportional to the number of nodes, i.e., the node count scales with the state-vector (e.g., 28-4, 30-16, 31-32, 32-64), the execution time increases with the number of nodes from 41.3 to 52.6~s. Outside this direct relationship (e.g., 24-1, 34-64, 36-64) the time is ruled by the number of amplitudes per node. On the capped 64-node allocation, each additional pair of qubits quadruples the work per node and the execution time (52.6, 235.1, and 1044.1~s for 32, 34, and 36 qubits).

Regarding reconfigurations, the fifth column of Table~\ref{tab:walltimes} reports the overhead times measured as in the previous section: from the decision until execution resumes.
The full overheads grow with the target allocation (1.41, 1.64, 2.14, and 2.02~s for the four expansions) and are within 4--14\% of those measured in isolation in Fig.~\ref{fig:reconf-matrix}.

\begin{table}[htbp]
\centering
\caption{Circuit execution times and transition reconfiguration overheads for the evaluated seven-circuit batch. Each transition takes place after the circuit in the same row; the first row is the initial single-node allocation.}
\label{tab:walltimes}
\setlength{\tabcolsep}{4pt}
\begin{tabular}{%
  >{\centering\arraybackslash}p{1.0cm}
  >{\centering\arraybackslash}p{1.0cm}
  >{\centering\arraybackslash}p{1.5cm}
  >{\centering\arraybackslash}p{1.5cm}
  >{\centering\arraybackslash}p{2.0cm}
}
\toprule
\textbf{Qubits} & \textbf{Nodes} & \textbf{Execution Time (s)} & \textbf{Transition} & \textbf{Reconfiguration Time (s)} \\
\midrule
-- & 1  & --       & $1 \rightarrow 4$  & 1.41 \\
28 & 4  & 41.331   & $4 \rightarrow 16$ & 1.64 \\
30 & 16 & 47.050   & $16 \rightarrow 1$ & 0.44 \\
24 & 1  & 7.944    & $1 \rightarrow 64$ & 2.14 \\
36 & 64 & 1044.113 & $64 \rightarrow 32$ & 0.81 \\
31 & 32 & 49.345 & $32 \rightarrow 64$ & 2.02 \\
32 & 64 & 52.593  & --                  & --    \\
34 & 64 & 235.057  & --                  & --    \\
\bottomrule
\end{tabular}
\end{table}

\section{Discussion}\label{sec:discussion}
In this section, we discuss the implications of our findings from the evaluation of the integration of QuEST with DMR. We analyze the benefits and limitations of using malleability for quantum circuit simulations on HPC systems, and we explore potential future directions for research in this area.

The current right-sizing policy keeps a fixed number of local qubits per node,
so the allocation grows with the size of the state-vector regardless of the
circuit. However, the optimal allocation from a performance standpoint depends
on the circuit itself, since communication overhead grows with the number of
global qubits and can outweigh the benefit of additional parallelism, especially for
circuits with a low computational load. Incorporating performance models or online profiling into the
right-sizing decision would allow the simulator to pick a better trade-off
between parallelism and communication cost~\cite{aréjulaaísa2026communicationawarequbitplacementautomatic}.

The execution model of the \texttt{circuit\_executor} also leaves room for improvement.
Currently, circuits run sequentially: the next circuit starts only when the
previous one has finished. A natural extension is to allow concurrent
execution. For example, if a 32-node circuit is running and three 16-node
circuits are waiting in the queue, the executor could request 16 additional
nodes in advance. Once the 32-node circuit finishes and those nodes are freed,
the three smaller circuits could start simultaneously using all 48 nodes at
once, instead of running one after another. This would improve resource
utilization and reduce the total time to complete a workload.


Variational algorithms alternate circuit evaluations with a classical optimizer that chooses the parameters of each evaluation from the results of the previous ones~\cite{cerezo_vqa_2021}. Within one problem instance, all evaluations use the same number of qubits, so the loop triggers no resize. Resizes appear when instances of increasing size are chained, as in the QAOA studies that transfer parameters from smaller to larger instances~\cite{galda_transfer_2023,shaydulin_labs_2024}, and then the optimizer state must survive each restart. The current executor already supports this case: since a circuit is a user-defined function, it can read its results from the register, run the optimizer update, and persist the new parameters before it returns, so that the checkpoint written after the circuit also covers this classical step. A natural extension is to let users register such state with the executor, which would then save it in the checkpoint next to the index of the next circuit and restore it transparently after a restart.

More generally, malleability currently applies to the circuits of the batch, not to the whole application: the code outside the executor runs again after every restart, so it should be limited to setting up the environment. Making complete hybrid workflows malleable, including their classical stages, is left for future work.

A further step is a persistent, adaptive simulation service. Instead of one job per workflow, a long-running service could hold a pool of nodes and run circuits as they arrive, resizing the pool for each of them. Circuits would then avoid the queue wait and the startup of a new job; only expansions of the pool would wait for the scheduler to grant nodes. CUNQA~\cite{vazquez_cunqa_2025} already deploys simulators as a service on HPC systems, where users submit circuits to a pool of virtual quantum processing units (vQPUs), but the resources and layout of these vQPUs are fixed when the service is deployed. Combined with DMR, such a service could change at each reconfiguration not only its number of nodes, but also how they are grouped into vQPUs and how each vQPU is configured, since every restart rebuilds the processes from scratch. The virtual hardware would then match the circuits being submitted.

\section{Conclusion}\label{sec:conclusion}
This paper has presented a malleable version of the QuEST simulator for sequences of circuits whose number of qubits changes from one circuit to the next. Built on the DMR framework, a single job adapts its number of nodes to each circuit, instead of holding the allocation of the largest circuit for the whole sequence or submitting one job per circuit.

The design reconfigures only between circuits, when no state-vector is alive. As a result, no amplitudes are redistributed, the simulation kernels of QuEST remain unchanged, and a checkpoint holding the index of the next circuit is enough to resume after a resize. A right-sizing policy maps the number of qubits of each circuit to a power-of-two number of nodes, and the \texttt{circuit\_executor} API hides the reconfigurations: users wrap and register their circuits, and can submit the job on a single node without sizing it in advance.

On MareNostrum~5, reconfigurations between 1 and 64 nodes measured in isolation take 0.34 to 0.78~s for shrinkages and 1.23 to 2.02~s for expansions, which must also acquire the new nodes and extend the PRRTE distributed virtual machine before spawning processes. 

Future work includes right-sizing policies based on performance models that balance parallelism against communication, concurrent execution of circuits within the executor, checkpoints that also store user-defined state such as optimizer parameters, and a persistent malleable simulation service for HPC-QC workflows.

\section*{Acknowledgment}
BSC researchers have been financially supported by the Ministry of Economic Affairs and Digital Transformation of the Spanish Government through the QUANTUM ENIA project call - Quantum Spain project, and by the Ministerio para la Transformación Digital y de la función pública, within the framework of the Plan de Recuperación Transformación y Resiliencia, and by the European Union --- NextGenerationEU and the framework of the Digital Spain 2026 Agenda.

Antonio J. Peña was partially supported by the Ramón y Cajal fellowship RYC2020-030054-I funded by MCIN\slash AEI\slash 10.13039\slash 501100011033 and by ``ESF Investing in your future''.

The views and opinions expressed are solely those of the authors and do not necessarily reflect those of the European Union. Neither the European Union nor the European Commission can be held responsible for them.


Language polishing was performed using an AI language model, with subsequent thorough review and editing by the authors. The authors take full responsibility for the final manuscript content.

\bibliographystyle{IEEEtran}
\bibliography{bib/refs}

\end{document}